\documentclass[conference]{IEEEtran}
\IEEEoverridecommandlockouts

\usepackage{cite}
\usepackage{amsmath,amssymb,amsfonts,mathtools,bm}
\usepackage{algorithm}
\usepackage{algpseudocode}
\usepackage{graphicx}
\usepackage[caption=false,font=footnotesize]{subfig}
\usepackage{stfloats}
\usepackage{float}
\usepackage{textcomp}
\usepackage{xcolor}
\usepackage{balance}
\usepackage{placeins}
\usepackage{booktabs}
\usepackage{multirow}
\usepackage{array}
\usepackage{siunitx}
\usepackage{url}
\usepackage{enumitem}
\usepackage{tikz}
\usetikzlibrary{arrows.meta,positioning,calc,decorations.pathreplacing}

\usepackage[]{hyperref}
\usepackage{hyperref}
\usepackage{cleveref}
\def\BibTeX{{\rm B\kern-.05em{\sc i\kern-.025em b}\kern-.08em
T\kern-.1667em\lower.7ex\hbox{E}\kern-.125emX}}

\newcommand{\CaptureTime}{\SI{45}{ms}}
\newcommand{\CarrierSCS}{\SI{15}{kHz}}
\newcommand{\NRGridPRB}{66}

\newcommand{\WifiMCS}{2}
\newcommand{\APEPLengthBytes}{1024}

\newcommand{\SICFilterLen}{32}
\newcommand{\SICTrainFraction}{0.30}

\title{Over-the-Air Successive Interference Cancellation for Efficient 5G NR and Wi-Fi Spectrum Reuse\thanks{This work was supported by the UKRI/EPSRC Prosperity Partnership in Secure Wireless Agile Networks (SWAN), Grant EP/T005572/1, and the EPSRC Hub in All Spectrum Connectivity: Successive Interference Cancellation for Dynamic Spectrum Access (SINATRA), Grant EP/Y037197/1}}

\author{%
\IEEEauthorblockN{Mir Lodro, Francesco Raimondo, Geoffrey S. Hilton, Mark A. Beach, and Andrew C. M. Austin}
\IEEEauthorblockA{%
Communication Systems and Networks Research Group\\
School of Electrical, Electronic and Mechanical Engineering\\
University of Bristol, Bristol, United Kingdom\\
mir.lodro@bristol.ac.uk}
}
\begin{document}
\maketitle

\begin{abstract}
An over-the-air (OTA) experimental evaluation of concurrent 5G New Radio (5G NR) and Wi-Fi transmission using successive interference cancellation (SIC) in a shielded-box environment is presented. A USRP is used as the receiver, which captures the composite waveform containing both air-interface signals and applies sample-domain SIC to suppress the dominant 5G-NR signal and recover Wi-Fi signal from the residual waveform. The framework reports error vector magnitude (EVM), bit error rate (BER), sample-domain cancellation depth, and channel-estimate suppression, and, at the representative \(18\) dB attenuation point, measures \(11.88\) dB cancellation depth and \(26.96\) dB 5G channel suppression. The proposed methodology provides a practical basis for assessing cross-technology coexistence and receiver-side interference suppression under controlled OTA conditions.
\end{abstract}

\section{Introduction}
Efficient spectrum reuse between 5G and Wi-Fi is increasingly important as both systems expand into shared and adjacent spectrum bands \cite{ghosh2026recent}. In unlicensed operation, 5G New Radio (NR-U) follows shared-spectrum channel-access procedures defined by 3GPP, including sensing-based access and energy-detection rules~\cite{3gpp37213_rel18}. Consequently, most existing 5G/Wi-Fi coexistence approaches are framed at the medium-access layer, where coexistence is managed through listen-before-talk (LBT), random backoff, contention control, and energy-detection tuning rather than through physical-layer separation of overlapping waveforms~\cite{frangulea2025nruwifi,kosekszott2021downlink,keshtiarast2025wifi6e}. A key limitation of such approaches is that coexistence performance depends strongly on the collision domain observed by each transmitter. In heterogeneous deployments, asymmetric sensing, hidden nodes, unequal received powers, and carrier-sense threshold selection can lead to repeated collisions, degraded fairness, and inefficient airtime use~\cite{frangulea2025nruwifi}. Recent work has also shown that successive interference cancellation can be exploited explicitly for spectrum sharing over unlicensed bands, thereby motivating practical over-the-air (OTA) evaluation of cross-technology cancellation frameworks \cite{guo2024exploiting}. Recent studies of NR-U and Wi-Fi coexistence in the 5~GHz and 6~GHz bands, therefore, continue to evaluate performance primarily in terms of throughput fairness, channel-access behavior, maximum channel occupancy time, and energy-detection configuration~\cite{kosekszott2021downlink}. These limitations motivate receiver-side signal processing methods that move beyond collision avoidance. If the dominant interfering waveform can be estimated and suppressed at the receiver, then concurrent 5G and Wi-Fi transmissions need not be separated only in time or frequency. Instead, the same spectral resource may be reused more aggressively, provided that the residual waveform remains decodable for the cochannel system. Successive interference cancellation has been widely studied in full-duplex and interference-limited wireless systems as a practical means of reconstructing and subtracting a dominant interference component~\cite{Sabharwal2014FD,Bharadia2013FD}. In this work, we investigate this idea through an OTA SIC framework for concurrent 5G NR and Wi-Fi transmission. The main contributions of this work are as follows. First, we present a controlled OTA coexistence testbed for concurrent 5G NR and Wi-Fi transmission in a shielded measurement environment. Second, we implement a sample-domain SIC receiver based on waveform alignment, complex-gain fitting, and finite impulse response (FIR) channel estimation to suppress the dominant 5G interference component. Third, we define an evaluation methodology that jointly reports 5G and Wi-Fi EVM, BER, cancellation depth, and channel-estimate suppression. Finally, we demonstrate, through OTA measurements, that receiver-side cancellation can suppress the dominant 5G component and enable recovery of the weaker Wi-Fi waveform from the residual signal. The remainder of the paper is organized as follows. Section II presents the signal model and the successive interference cancellation method. Section III describes the experimental setup, including waveform configuration and receiver processing chain. Section IV defines the key performance metrics. Section V reports the experimental results and discussion. Section VI concludes the paper.
\section{System Model and SIC Method}
We consider a deployment in which an outdoor 5G transmission is reused by an indoor Wi-Fi system, such that the 5G signal is received as the dominant interferer while the indoor Wi-Fi signal is comparatively weaker at the receiver. The purpose of the cancellation stage is to suppress the dominant 5G component of the composite received waveform, allowing the weaker Wi-Fi signal to be recovered and evaluated more reliably.
\begin{figure}[t]
    \centering
    \includegraphics[width=0.95\linewidth]{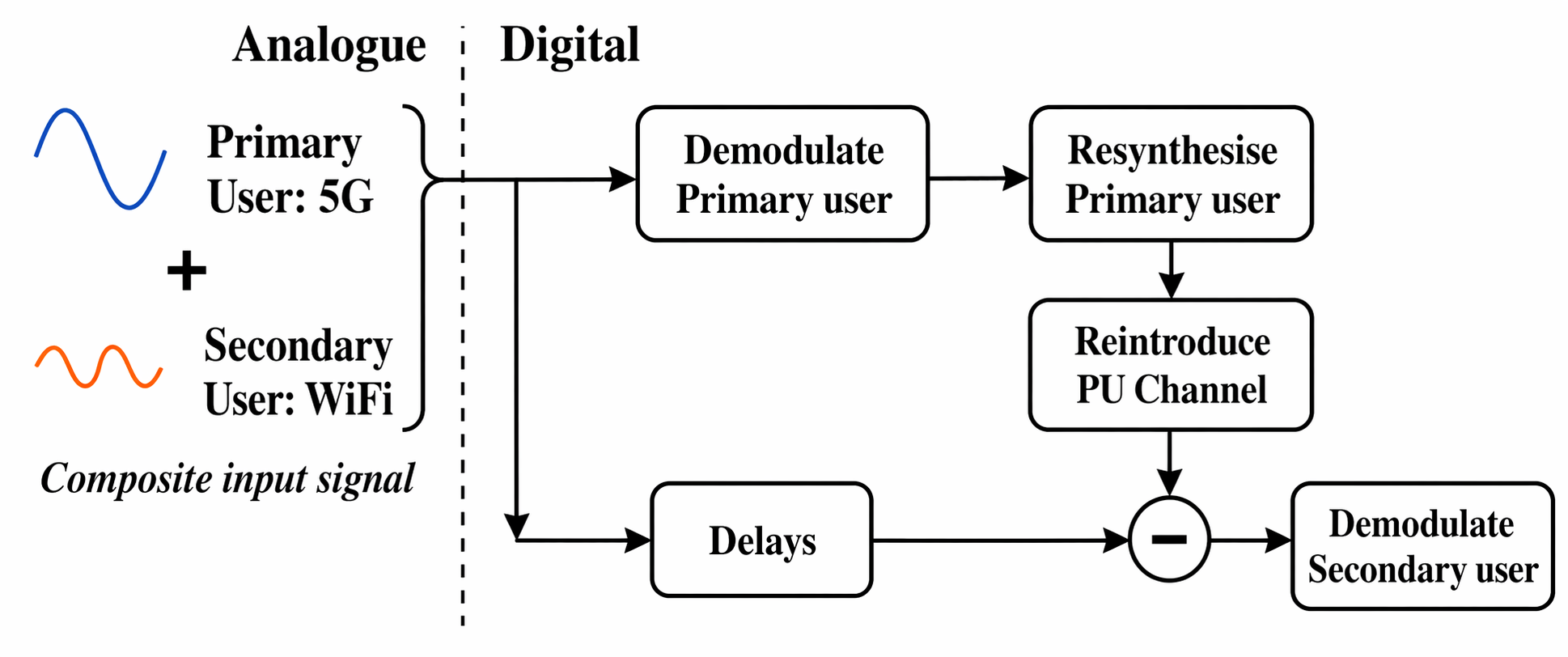}
    \caption{Conceptual signal flow of the proposed successive interference cancellation receiver.}
    \label{fig:sic_concept}
\end{figure}
Fig.~\ref{fig:sic_concept} illustrates the conceptual signal flow of the proposed SIC receiver. The composite received waveform contains a dominant 5G-NR component and a weaker Wi-Fi component. The receiver first demodulates and reconstructs the stronger 5G signal, then applies delay alignment, complex-gain correction, and channel reconstruction before subtracting the resulting estimate from the composite waveform. This sample-domain cancellation approach follows the general principle of digital interference reconstruction used in full-duplex and interference-cancellation receivers~\cite{Sabharwal2014FD,Bharadia2013FD}. Ideally, the residual waveform contains the Wi-Fi signal together with receiver noise and residual modelling error, allowing the secondary waveform to be decoded after suppression of the dominant 5G interference.
\subsection{Composite signal model}
Let $x_{\text{NR}}[n]$ denote the 5G NR baseband signal and $x_{\text{W}}[n]$ the Wi-Fi baseband signal transmitted by a second radio. The received composite waveform may be written as
\begin{equation}
    y[n] = h_{\text{NR}}[n] * x_{\text{NR}}[n] + h_{\text{W}}[n] * x_{\text{W}}[n] + w[n],
    \label{eq:rx_model}
\end{equation}
where $h_{\text{NR}}[n]$ denotes the effective channel seen by the 5G-NR signal, $h_{\text{W}}[n]$ denotes the Wi-Fi propagation channel, and $w[n]$ models receiver noise and residual impairments. The SIC receiver reconstructs an estimate of the dominant 5G-NR component as
\begin{equation}
    \hat{y}_{\text{NR}}[n] = \hat{h}_{\text{NR}}[n] * \bigl(\hat{\alpha}\,x^{\text{align}}_{\text{NR}}[n]\bigr),
    \label{eq:si_hat}
\end{equation}
where $x^{\text{align}}_{\text{NR}}[n]$ is the repeated 5G reference aligned to the captured waveform, $\hat{\alpha}$ is a complex scalar amplitude/phase correction, and $\hat{h}_{\text{NR}}[n]$ is an FIR estimate of the residual coupling response. The post-cancellation waveform is then
\begin{equation}
    y_{\text{res}}[n] = y[n] - \hat{y}_{\text{NR}}[n].
    \label{eq:residual_model}
\end{equation}
This reconstructed-subtraction model is standard in digital interference-cancellation architectures~\cite{Sabharwal2014FD,Bharadia2013FD}. Ideally, the residual waveform contains only the Wi-Fi signal and noise
\begin{equation}
    y_{\text{res}}[n]=x_{\text{W}}[n]+w[n]
\end{equation}
After coarse alignment, the receiver estimates a scalar least-squares correction
\begin{equation}
    \hat{\alpha} = \frac{\mathbf{x}^{H}\mathbf{y}}{\mathbf{x}^{H}\mathbf{x} + \epsilon},
    \label{eq:alpha_ls}
\end{equation}
followed by an $L_h$-tap regularized FIR fit. If $\mathbf{X}$ is the Toeplitz regression matrix formed from the aligned and scaled reference, and $\mathbf{y}$ is the training segment of the capture, then
\begin{equation}
    \hat{\mathbf{h}}_{\text{NR}} = \left(\mathbf{X}^{H}\mathbf{X} + \lambda \mathbf{I}\right)^{-1}\mathbf{X}^{H}\mathbf{y},
    \label{eq:firls}
\end{equation}
where $\lambda$ is a small regularization factor. This estimate is then used to synthesize the cancellation waveform across the full captured frame.
\section{Experimental Setup}
\subsection{Waveform Configuration}
The NR waveform uses normal cyclic-prefix OFDM with subcarrier spacing 15~kHz and resource allocation spanning all 66 physical resource blocks (PRBs). The physical downlink shared channel (PDSCH) occupies the full slot and uses 64-QAM modulation with one layer. Demodulation reference symbols (DM-RS) and phase-tracking reference symbols (PTRS) are enabled to support channel estimation and common-phase-error compensation. The waveform is generated for 2 frames, OFDM-modulated at its native NR sample rate 15.36~MSPS, normalized, and resampled to the common USRP rate of 20~MSPS for OTA transmission. The Wi-Fi waveform is generated using an IEEE 802.11ac very-high-throughput (VHT) configuration with channel bandwidth 20~MHz, MCS 2, payload length 1024 bytes, and 4 packets per generated burst. Idle spacing of 20~$\mu s$ is inserted between packets. The waveform is normalized and scaled before transmission to control the relative power in the composite channel. Quasi-static baseband channel models are also applied before transmission. For NR, a tapped-delay-line (TDL-B) and TDL-C profile with a delay spread of 50~$ns$ is used. For Wi-Fi, a TGn Model-B and Model-D profile is used with near-static fading. Although the OTA experiment already captures practical RF coupling, this pre-filtering stage enables controlled channel diversity while preserving the hardware signal flow.
\subsection{Hardware architecture}
The testbed uses USRPs operating at a common RF center frequency of 2.45~GHz. USRP~1 continuously transmits the 5G NR waveform and simultaneously captures the composite received signal, while USRP~2 transmits the Wi-Fi waveform. Both radios operate at a common baseband sample rate of 20~MSPS. Each experiment captures 45~$ms$ of complex baseband data, after which the mean DC component is removed. The captured waveform is then processed to evaluate 5G performance on the composite waveform and Wi-Fi performance on the post-SIC residual waveform.
\begin{figure}[t]
    \centering
    \includegraphics[width=\linewidth]{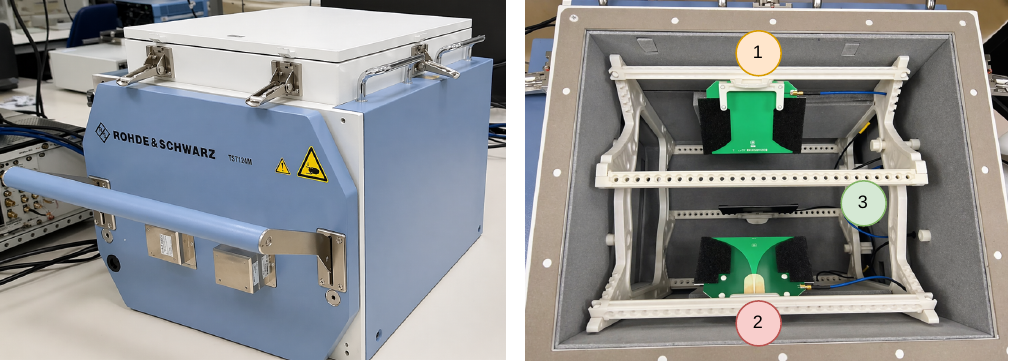}
    \caption{Photograph of the OTA measurement enclosure and antenna fixture used for the controlled SIC coexistence experiment. The arrangement provides a repeatable geometry for evaluating cancellation depth, residual interference, and cross-technology recovery performance. Labels 1, 2, and 3 are used for the wideband Vivaldi antennas.}
    \label{fig:ota_fixture}
\end{figure}
Fig.~\ref{fig:ota_fixture} shows the OTA measurement setup inside a high-performance RF shielding box R\&S TS7124M. Three broadband Vivaldi antennas with high transmission efficiency are mounted on three rails inside the chamber and connected to the NI USRP-2953R radios through the rear shielded RF feedthroughs using low-loss RF cables. A precision-calibrated RF attenuator (Weinschel Engineering Model~974-C1-4-72) with attenuation range 6--120~dB is inserted in the Wi-Fi transmit path to control the received Wi-Fi signal level. By sweeping the attenuation over a range of values, the relative strength of the Wi-Fi signal with respect to the cochannel 5G-NR signal can be varied in a controlled and repeatable manner. At each attenuation setting, the EVM and BER performance of both 5G-NR and Wi-Fi are measured. In addition, the sample-domain cancellation depth and the suppression of the estimated 5G channel are evaluated to quantify the effectiveness of the proposed SIC receiver. This arrangement provides a controlled and repeatable radiated geometry for concurrent 5G-NR and Wi-Fi transmission while maintaining isolation from the external RF environment. The shielding box provides shielding efficiency of at least 80 dB and therefore enables consistent evaluation of cancellation depth, residual interference, and cross-technology recovery performance under OTA conditions.

\subsection{Receiver processing chain}
The receiver processing flow is as follows. First, the composite waveform is captured while 5G and Wi-Fi are both active. The composite waveform is then resampled from the USRP rate to the native NR rate for 5G demodulation. Timing is estimated, the OFDM symbols are demodulated, DM-RS-based channel estimation is performed, the PDSCH symbols are equalized, and 5G EVM and alignment-aware BER are computed. Next, the interference delay is estimated by cross-correlating the repeated NR reference with the capture. A scalar coefficient and FIR SIC channel are then fitted using the training portion of the capture. The reconstructed 5G estimate is subtracted to form the residual waveform. The Wi-Fi receiver operates directly on the USRP sample rate. Finally, Wi-Fi packet detection, carrier frequency offset (CFO) estimation and correction, timing synchronization, channel estimation, VHT field recovery, and data decoding are applied to the residual waveform. The receiver logs field-level Wi-Fi EVM metrics, including L-SIG, SIG-A, SIG-B, and data-field EVM, together with coarse and fine carrier-frequency-offset estimates and timing offsets. The main waveform and experimental parameters are summarized in Table~\ref{tab:params}.
\begin{table}[t]
    \centering
    \caption{Waveform and experimental parameters.}
    \label{tab:params}
    \begin{tabular}{ll}
        \toprule
        Parameter & Value \\
        \midrule
        RF center frequency & 2.45~GHz \\
        USRP sample rate & 20~MSPS \\
        Capture time & \CaptureTime{} \\
        5G NR subcarrier spacing & \CarrierSCS{} \\
        5G NR grid size & \NRGridPRB{} PRBs \\
        5G modulation & 64-QAM PDSCH \\
        Wi-Fi mode & IEEE 802.11ac VHT \\
        Wi-Fi bandwidth & 20~MHz \\
        Wi-Fi MCS & \WifiMCS{} \\
        Wi-Fi payload length & \APEPLengthBytes{} bytes \\
        SIC filter length & \SICFilterLen{} taps \\
        SIC training fraction & \SICTrainFraction{} \\
        \bottomrule
    \end{tabular}
\end{table}
\section{Performance Metrics}
\subsection{5G and Wi-Fi quality metrics}
The 5G receiver reports RMS EVM over equalized PDSCH symbols and the BER. EVM is a standard modulation-quality metric used in both NR and IEEE 802.11 systems~\cite{3gpp38104_rel18,ieee80211_2020}. Let $s_k$ denote the equalized constellation samples and $\hat{s}_k$ the nearest reference symbols. The RMS EVM is reported as
\begin{equation}
    \mathrm{EVM}_{\mathrm{RMS}}(\%) = 100\sqrt{\frac{\sum_k |s_k - \hat{s}_k|^2}{\sum_k |\hat{s}_k|^2}}.
    \label{eq:evm}
\end{equation}
For Wi-Fi, the receiver reports field-level EVM and packet BER after VHT recovery. The main Wi-Fi summary metrics used in this paper are the mean VHT-Data EVM and the mean physical service data unit (PSDU) BER across decoded packets. These metrics are evaluated on both the composite waveform and the post-SIC residual waveform.

\subsection{SIC depth and channel suppression}
The sample-domain SIC depth is computed from the average received power before and after cancellation:
\begin{equation}
    D_{\mathrm{SIC}} = 10\log_{10}\left(\frac{P_{\mathrm{before}}}{P_{\mathrm{after}}}\right)\,\text{dB}.
    \label{eq:sic_depth}
\end{equation}
where \(P_{\mathrm{before}}\) is the average received power of the composite waveform before cancellation and \(P_{\mathrm{after}}\) is the average received power after cancellation. Thus, \(D_{\mathrm{SIC}}\) measures the power reduction achieved by SIC in the sample domain. In addition, the implementation reports a channel-suppression metric based on the norm ratio between the 5G channel-estimate grid before and after SIC:
\begin{equation}
    S_{H} = 20\log_{10}\left(\frac{\lVert \mathbf{H}_{\mathrm{before}} \rVert_2}{\lVert \mathbf{H}_{\mathrm{after}} \rVert_2}\right)\,\text{dB}.
    \label{eq:chan_supp}
\end{equation}
Where \(\mathbf{H}_{\mathrm{before}}\) and \(\mathbf{H}_{\mathrm{after}}\) are the estimated 5G channel-response grids before and after cancellation, respectively, \(\|\cdot\|_2\) denotes the matrix \(2\)-norm, and \(S_H\) quantifies the reduction in the coherent 5G channel structure after cancellation.
\section{Results and Discussion}
At the representative \(18\) dB attenuation operating point, the composite waveform is dominated by the 5G-NR signal, enabling 5G demodulation while Wi-Fi decoding on the composite waveform remains unsuccessful. After SIC, the reconstructed 5G component is suppressed and the weaker Wi-Fi waveform becomes decodable on the residual signal. This representative operating point therefore illustrates the intended SIC behavior: degradation of the cancellable 5G component together with improved recoverability of the cochannel Wi-Fi transmission after subtraction of the dominant interference.

\subsection{Cancellation depth and channel suppression}
The measured sample-domain SIC depth reflects aggregate power reduction in the received waveform, whereas channel suppression quantifies the reduction of coherent 5G channel structure after cancellation. Both metrics are important because a reduction in received power alone does not guarantee that the residual waveform has become sufficiently clean for Wi-Fi recovery.
\begin{figure}[t]
    \centering
    \includegraphics[width=0.95\linewidth]{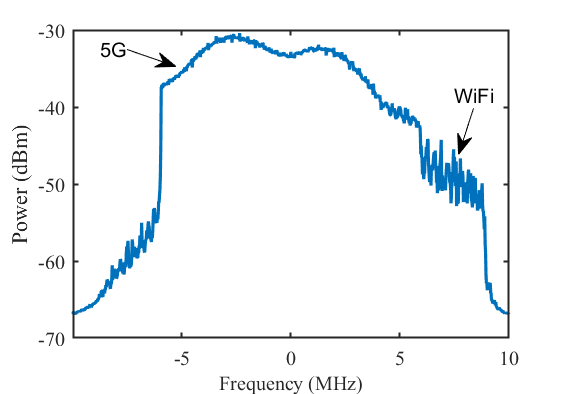}
    \caption{Measured spectrum of the received composite waveform. The plot confirms the occupied bandwidth around the shared center frequency and provides a compact view of the dominant in-band energy before digital cancellation.}
    \label{fig:rxpower}
\end{figure}
Fig.~\ref{fig:rxpower} shows the measured power spectrum of the composite received waveform before SIC. The occupied in-band region confirms concurrent 5G-NR and Wi-Fi transmission in the shared channel. These results indicate that the dominant spectral energy in the composite capture is largely due to the cancellable 5G-NR component and help explain why Wi-Fi decoding becomes possible only after cancellation.

\begin{figure}[t]
    \centering
    \includegraphics[width=\linewidth]{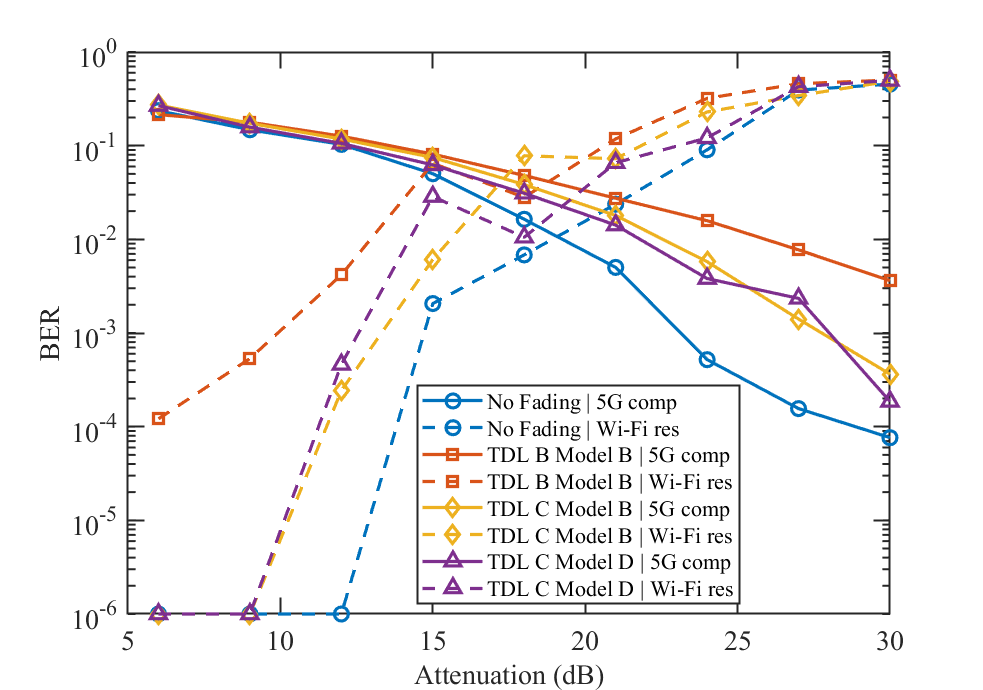}
    \caption{Measured BER versus attenuation for the 5G signal on the composite waveform and the Wi-Fi signal on the residual waveform under multiple channel conditions.}
    \label{fig:ber_attn}
\end{figure}
\begin{figure}[t]
    \centering
    \includegraphics[width=\linewidth]{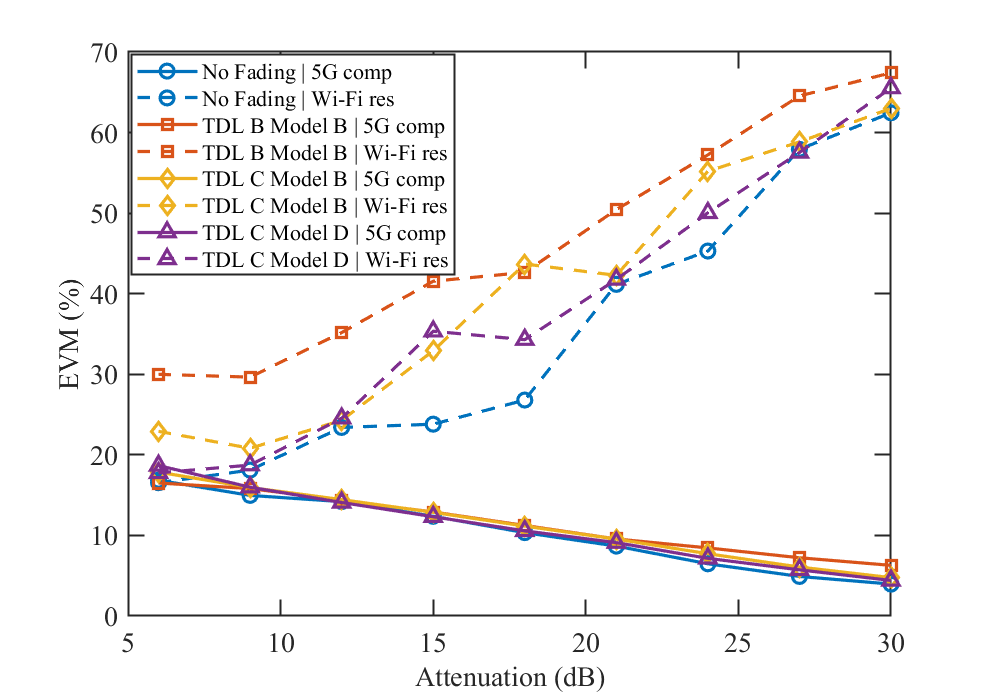}
    \caption{Measured EVM versus attenuation for the 5G signal on the composite waveform and the Wi-Fi signal on the residual waveform under multiple channel conditions.}
    \label{fig:evm_attn}
\end{figure}
Fig.~\ref{fig:ber_attn} and Fig.~\ref{fig:evm_attn} show the BER and EVM performance, respectively, under a controlled attenuation sweep applied in the Wi-Fi transmit path using a precision calibrated attenuator. As the attenuation increases, the received Wi-Fi signal is reduced relative to the cochannel 5G-NR signal. Fig.~\ref{fig:ber_attn} shows the corresponding BER trends for the 5G-NR signal on the composite waveform and the Wi-Fi signal on the post-SIC residual waveform. Fig.~\ref{fig:ber_attn} and Fig.~\ref{fig:evm_attn} show the attenuation-sweep behavior of the 5G and Wi-Fi links under the no-fading and filtered channel cases. In these plots, the 5G metric is the RMS EVM or BER measured on the composite waveform, whereas the Wi-Fi metric is the mean VHT-Data EVM or mean BER measured on the post-SIC residual waveform. Across all four channel cases, Wi-Fi decoding fails on the composite waveform but succeeds on the residual waveform after cancellation. The no-fading case gives the strongest residual Wi-Fi recovery, while the filtered cases exhibit higher residual EVM and BER, indicating that channel conditions affect the recovery margin after SIC.
To provide a clearer cross-condition comparison, Table~\ref{tab:sic_quality} summarizes the main 5G composite and Wi-Fi residual key performance indicators (KPIs) at the representative attenuation of \(18\) dB.
\begin{table}[t]
\caption{5G composite and Wi-Fi residual KPIs at 18 dB attenuation.}
\label{tab:sic_quality}
\centering
\scriptsize
\setlength{\tabcolsep}{4pt}
\begin{tabular}{lcccc}
\hline
Case & 5G EVM & 5G BER & Wi-Fi EVM & Wi-Fi BER \\
 & (\%) & comp. & (\%) & res. \\
\hline
No fading       & 10.35 & $2\times10^{-2}$ & 26.80 & $7\times10^{-3}$ \\
TDL-B / Model-B & 11.23 & $5\times10^{-2}$ & 42.65 & $3\times10^{-2}$ \\
TDL-C / Model-B & 11.15 & $4\times10^{-2}$ & 43.69 & $8\times10^{-2}$ \\
TDL-C / Model-D & 10.57 & $3\times10^{-2}$ & 34.30 & $1\times10^{-2}$ \\
\hline
\end{tabular}
\end{table}
As shown in Table~\ref{tab:sic_quality}, the no-fading case provides the best residual Wi-Fi recovery, with the lowest residual EVM and BER. In contrast, the filtered channel cases exhibit higher residual Wi-Fi distortion and error rate, even though the 5G composite metrics remain within a relatively narrow range. This indicates that post-SIC Wi-Fi recovery is more sensitive to channel conditions than the pre-SIC 5G composite quality.
\begin{figure}[t]
    \centering
    \includegraphics[width=0.95\columnwidth]{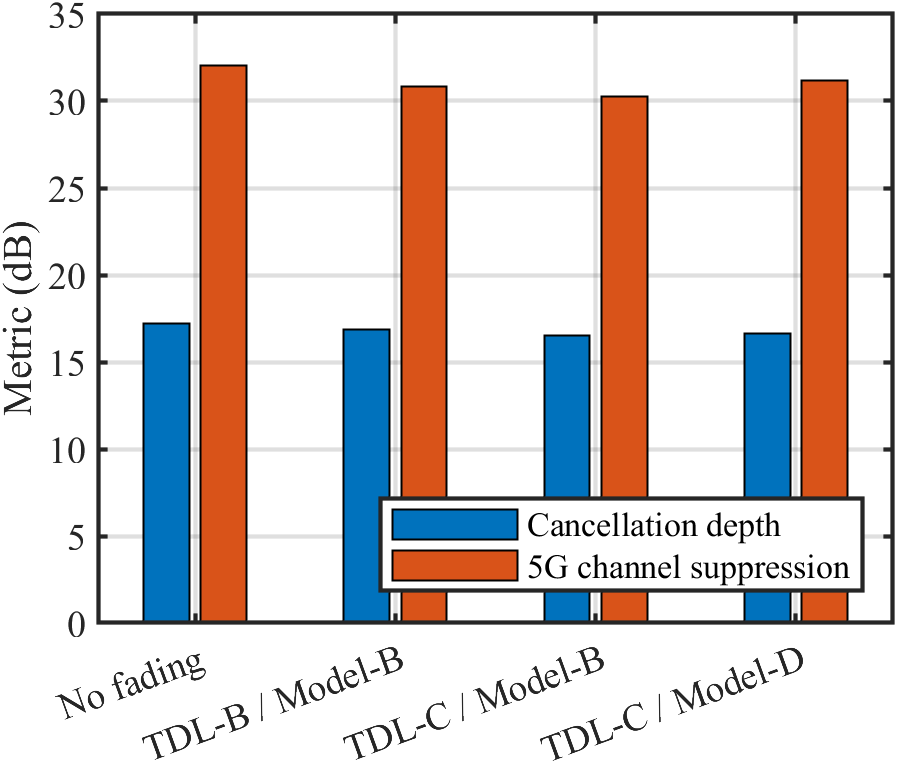}
    \caption{Comparison of sample-domain cancellation depth and 5G channel suppression across the evaluated channel cases at \(18\) dB attenuation.}
    \label{fig:bar_sic_18db}
\end{figure}
\begin{figure}[!t]
    \centering
    \includegraphics[width=0.95\columnwidth]{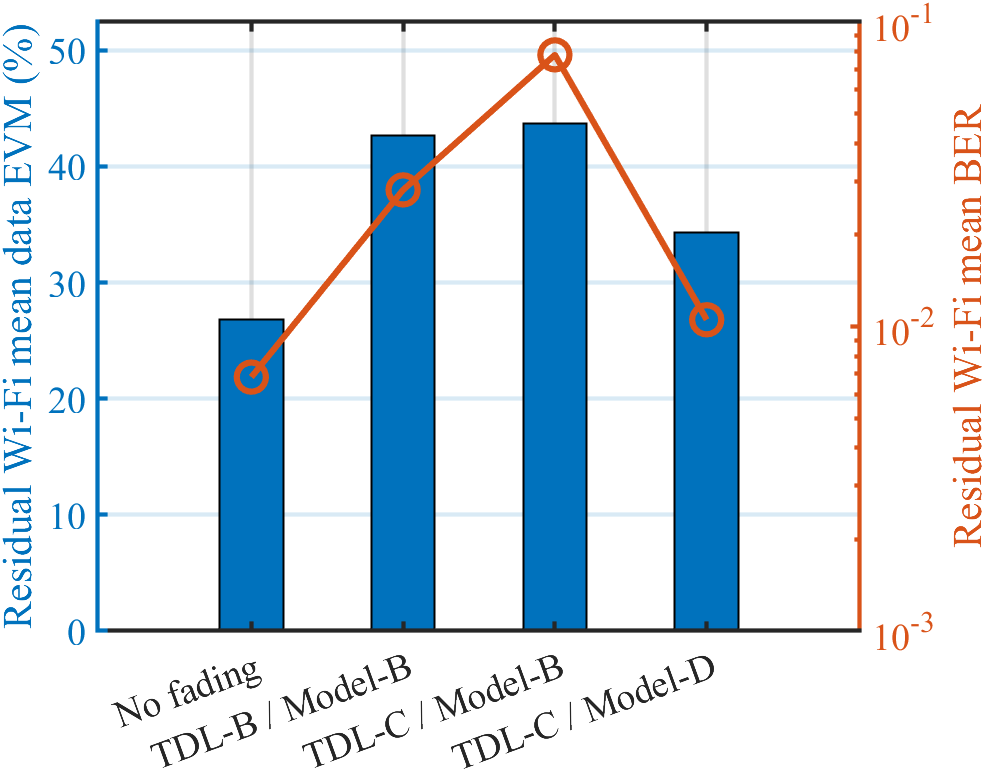}
    \caption{Comparison of residual Wi-Fi mean data EVM and mean BER across the evaluated channel cases at \(18\) dB attenuation after successive interference cancellation.}
    \label{fig:bar_wifi_18db}
\end{figure}
To provide a clearer cross-condition comparison, Fig.~\ref{fig:bar_sic_18db} and Fig.\ref{fig:bar_wifi_18db} summarize the results at a representative attenuation of \(18\) dB. Fig.~\ref{fig:bar_sic_18db} shows that the measured cancellation depth and 5G channel suppression vary only moderately across the evaluated channel cases, indicating broadly consistent suppression of the dominant 5G component. In contrast, Fig.~\ref{fig:bar_wifi_18db} shows larger variation in the residual Wi-Fi EVM and BER, with the no-fading case giving the strongest recovery and the filtered channel cases showing degraded residual quality. These results indicate that post-SIC Wi-Fi recovery depends not only on gross cancellation depth but also on the residual error structure induced by the channel condition.
\section{Conclusion}
This paper has presented an OTA experimental framework for evaluating successive interference cancellation between concurrent 5G NR and Wi-Fi transmissions in a controlled shielded-box environment. The proposed receiver reconstructs and subtracts the dominant 5G-NR waveform in the sample domain using alignment, complex-gain fitting, and FIR channel estimation. At the representative 18 dB attenuation operating point, the measured cross-condition results show residual Wi-Fi EVM ranging from 26.80\% to 43.69\% and residual Wi-Fi BER ranging from \(7\times10^{-3}\) to \(8\times10^{-2}\), while the corresponding 5G composite EVM remains within a relatively narrow range of 10.35\% to 11.23\%. The resulting framework jointly reports 5G and Wi-Fi EVM, BER, cancellation depth, and channel-estimate suppression. These results show that receiver-side cancellation can suppress the dominant 5G interference sufficiently to enable recovery of the weaker cochannel Wi-Fi transmission.


\balance
\bibliographystyle{IEEEtran}
\bibliography{references}

\end{document}